# Temporal evolution of mesoscopic structure of some non-Euclidean systems using a Monte Carlo model

T. Mazumdar[1], S. Mazumder[2*] and D. Sen[2]

[1]*Research Reactor Services Division, Bhabha Atomic Research Centre, Mumbai-400 085, India*

[2]*Solid State Physics Division, Bhabha Atomic Research Centre, Mumbai-400 085, India*

* To whom correspondence should be addressed, E-mail: smazu@barc.gov.in

**Abstract**

*A Monte Carlo based computer model is presented to comprehend the contrasting observations of Mazumder et al. [Phys. Rev. Lett.* ***93****, 255704 (2004) and Phys. Rev. B* ***72****, 224208 (2005)], based on neutron-scattering measurements, on temporal evolution of effective fractal dimension and characteristic length for hydration of cement with light and heavy water. In this context, a theoretical model is also proposed to elucidate the same.*

## 1. Introduction:

Dispersion of liquids in solids leading to gel formation is a diffusion controlled process. Diffusion in Euclidean geometry is formulated theoretically by random walk of drunken walker in Euclidean space. Gelation of silicates, cementitious material, leads to mesoscopic structures with non-Euclidean fractal morphology in the length scale of order $10–10^4$ $A^0$. Diffusion in non-Euclidean fractal geometry is similar, in physical terms, to random walk of drunken walker on a road system designed by another drunken Engineer–theoretical formations of which are not satisfactorily comprehensible[1,2] yet like dynamics of many hydration reactions. It is shocking that investigation of cement is still in its infancy despite the fact that cement is ubiquitous material which is indispensable in the construction industry, in nuclear energy programs for immobilization of non heat-generating low-level radioactive waste, and in the petroleum industry to line oil wells by pumping cement slurry to isolate productive zones and with global production exceeding that of any other material of technological importance. The total world consumption of cement in 2008 was about 2.5 billion metric tons almost double of that of steel. Manufacturing of cement contributes about 4% of global and 5-7% of the total man-made $CO_2$ emissions. The understanding of the mechanism of its hydration and evolution of cement-water mixtures into a material of high compressive strength is paramount to improve its life time and other macroscopic properties such as compressive strength, permeability, elastic modulus etc.

Cement reacts with light water via hydration reaction which yields an amorphous calcium-silicate-hydrate (C-S-H) gel-like structure and crystalline calcium hydroxide as the main products. To elucidate the microscopic structure of the C-S-H gel, many models[3-8] have been proposed. Investigations, based on small-angle neutron scattering (SANS), on continuous temporal evolution of mesoscopic structure during hydration of cement are recent[9-11]. SANS measurements involve mapping of time-dependent scattering function $S(q, t)$ where t stands for time and $q$ is the modulus of the scattering vector $\boldsymbol{q}$. Because of isotropic nature of the system at the mesoscopic scale, $S(q,t)$ is a function only of $q$. For all the experimental measurements of concern, it has been observed that $S(q,t)$ asymptotically approaches a form $S(q,t) \sim q^{-\eta(t)}$. The exponent η, associated with power-law scattering, reflects[12] directly the mass fractal dimension $D_m$. For a mass fractal object, $\eta = D_m$ with $1 < \eta < 3$ and $1 < D_m < 3$. For objects whose surface is fractal, the exponent η is related[13] to the surface fractal dimension $D_s$ with $3 < \eta < 4$ and $2 < D_s < 3$.

The investigations[2,9-11,14] on hydration of silicates and sulphates with light water ($H_2O$) and heavy water ($D_2O$) report some seemingly incomprehensible results. It is observed that kinetics of hydration of silicates and sulphates are of non-linear nature even at the initial time. In case of hydration of silicates with light water, the hydrating mass exhibits mass fractal nature throughout the hydration, with the mass fractal dimension increasing with time and reaching a finite saturation value at large time. The second phase grows with time initially. Subsequently, the domain size of the second phase saturates. It has also been demonstrated that light water hydration of silicates exhibits a scaling phenomenon for a characteristic length $L(t)$ with a measure of curvature of normalized[2] time-dependent scattering function $S(q, t)$.

For a monodisperse population of spheres of radius $R$, $L(t)=R\sqrt{2/5}$. For a polydisperse population of spheres, with number density $\rho(R)$ of radius R, $L(t)=\sqrt{2\langle R^8\rangle/5\langle R^6\rangle}$ where $\langle R^n\rangle$ is the $n$-th moment of the distribution $\rho(R)$. The temporal behavior of the characteristic length has been observed to be far from a power law. Further, temporal evolution of $\{L(t)\}^2$ nearly mimics the trend of evolution of $D_m$. $D_m$ also reaches a plateau almost at the same time (Fig.1 for hydration of tricalcium silicate with light water to cement ratio of 30%). It is interesting to note that these observations indicate that both physical quantities, characteristic length and effective fractal dimension in the length scale (it is basically effective fractal dimension for the length scales from $10^3$-$10^4 A^0$; for simplicity we call it fractal dimension everywhere in the text), having different dimensionality (as fractal dimension is a dimensionless quantity while characteristic length has a dimension of length) reach a plateau almost at the similar time and almost in a similar fashion. We want to understand whether the similarity of temporal evolutions of altogether two different physical quantities is an accidental one or not. But repeated measurements[2], varying over wide range of compositions, brought out this phenomenal accidental similarity.

As far as chemistry is concerned, the hydration of silicates with light and heavy water is expected to be quite similar except for kinetics. Due to the different molecular masses of light and heavy water, diffusion is expected to be more sluggish for heavy water. Further, it is known that hydrogen bond with deuterium is slightly stronger than the one involving ordinary hydrogen[15]. The lifetime of hydrogen bond involving D is longer vis-à-vis that involving H because the libration motions perpendicular to the bond direction have smaller amplitude for D than that for H, because of difference of masses. If diffusion is the only controlling factor, the hydration of cement with $H_2O$ and $D_2O$ is expected to be quite similar except for kinetics as

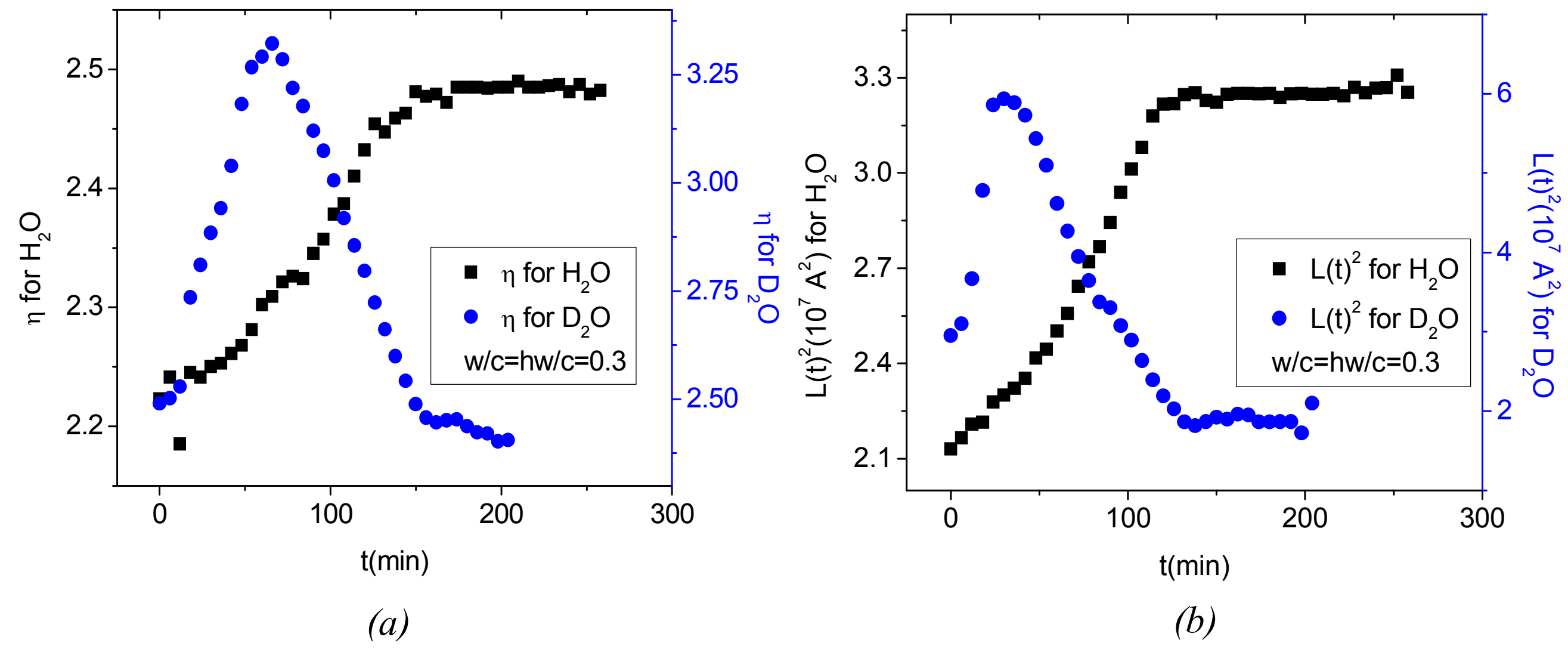

**Fig.1.** *(a) Temporal Evolution of $\eta$ for both light water and heavy water hydrating silicates as observed from neutron scattering measurements*[1]*. Only a typical representative figure has been considered. (b) Temporal Evolution of $(L(t))^2$ for both light water and heavy water hydrating silicates as observed from neutron scattering measurements*[1]*. Only a typical representative figure has been considered. For light water hydrating silicates, both $\eta$ and $(L(t))^2$ increase with*

*time and reach a plateau whereas for heavy water hydrating silicates, they first increase, then reach a peak and finally decrease as time approaches.*

diffusion of $D_2O$ is somewhat sluggish because of its heavier molecular mass vis-à-vis $H_2O$.

However, some incomprehensible contrasting behavior has been observed in the case of hydration of silicates with heavy water as far as the kinetics of new phase formation is concerned (Fig.1 for hydration of tricalcium silicate with heavy water to cement ratio of 30%).The domain size of the density fluctuations grows in the beginning for a while, and subsequently, appears to decrease with time, reaching saturation ultimately. In the case of hydration of silicates with

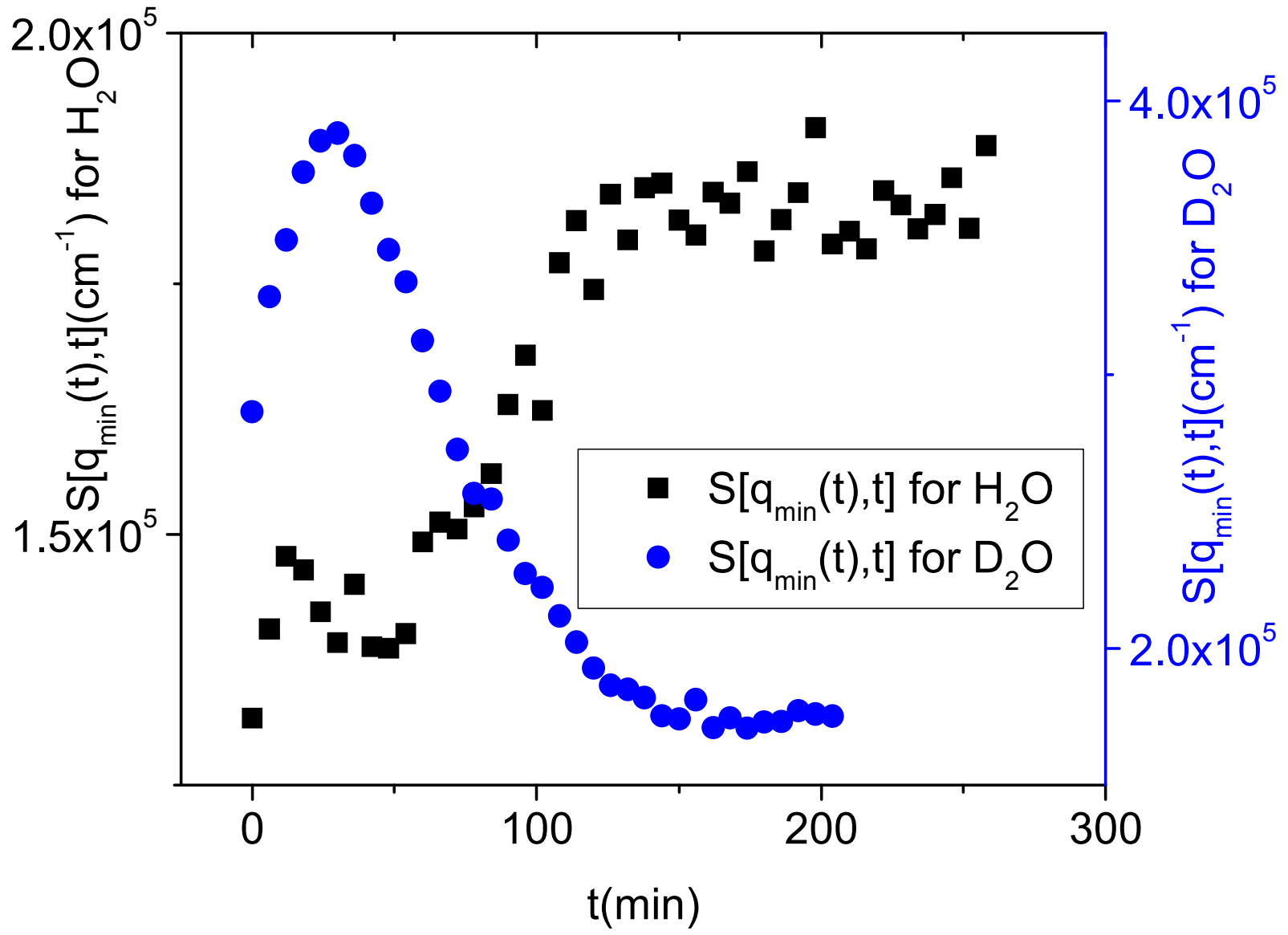

***Fig.2.*** *Temporal Evolution of $S(q_{min}(t),t)$ for light water and heavy water*[1] *hydrating silicates where $S(q,t)$ represents scattering function and $q_{min}$ is the lowest attained q value for a particular measurement. It resembles the nature of temporal evolution of $\eta$ and $(L(t))^2$ for light and heavy water hydration(Fig.1). We have considered only a typical representative figure here.*

heavy water, the microstructure of the hydrating mass undergoes a transition from mass fractal to surface fractal and subsequently to mass fractal. The scaling phenomenon, with all possible measures of the characteristic length, has not been established for the hydration of silicates with heavy water. It is a conjecture that the different rates of diffusion of light and heavy water in forming a gel structure in silicates lead to the formation of different structural networks with different scattering contrasts. (Fig.2)

In recent past, a Monte Carlo simulation, for embedded Euclidean dimension of 2, has been attempted to explain aforementioned incomprehensible observations[16]. This simulation could explain a few of the experimental results only for light water hydrating cement paste, by employing the concept of deposition mechanism of C-S-H gel in the available spaces inside the fractal cluster. However, the contrasting features of the structural evolution when hydrated with heavy water could not be explained on the same footings. Further, the earlier simulation was performed in embedded Euclidean dimension of 2 due to limitation in computing resources.

In the present paper, we proposed a computer model that elucidates the contrasting experimental observations during hydration of cement both by light and heavy water. Present simulation, unlike the earlier one, has also been extended to three dimension (3D). Further, a theoretical model has been proposed to explain only the temporal evolutions of fractal dimension for both light water and heavy water hydration cases.

## 2. Computer model:

In the present section, a Monte Carlo based computer model will be described to elucidate the hydration reaction of cement particles with light water as well as heavy water. Temporal evolutions of fractal dimension and characteristic length of the system during hydration will be dealt with both in two and three dimensions.

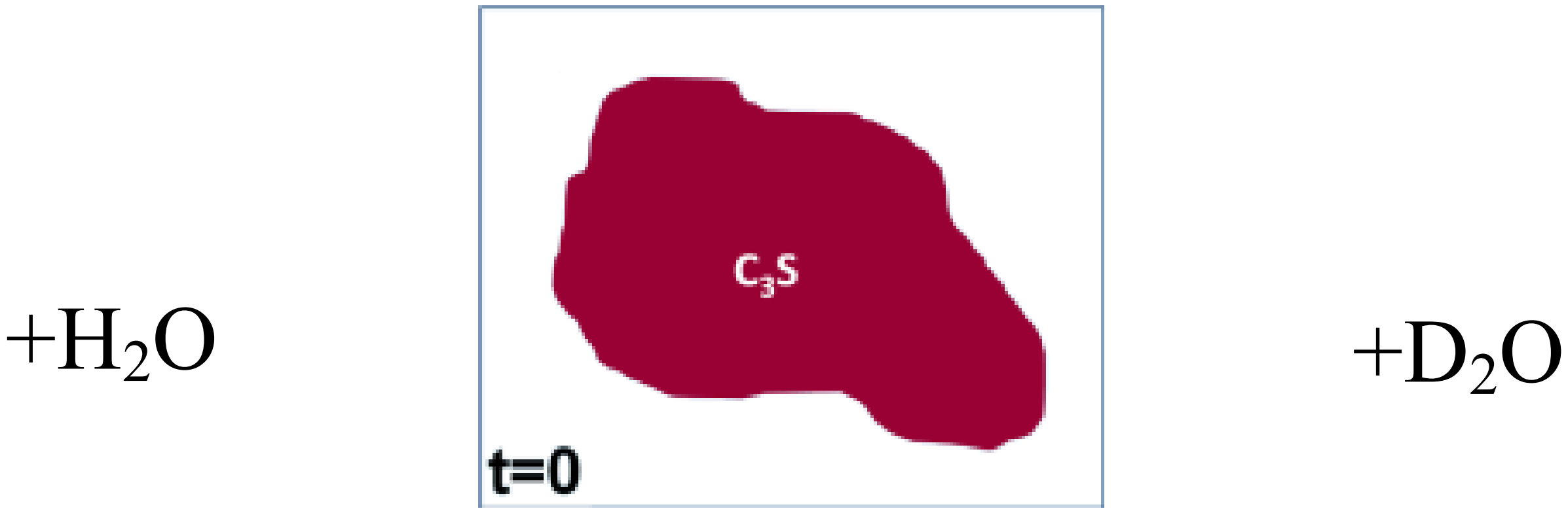

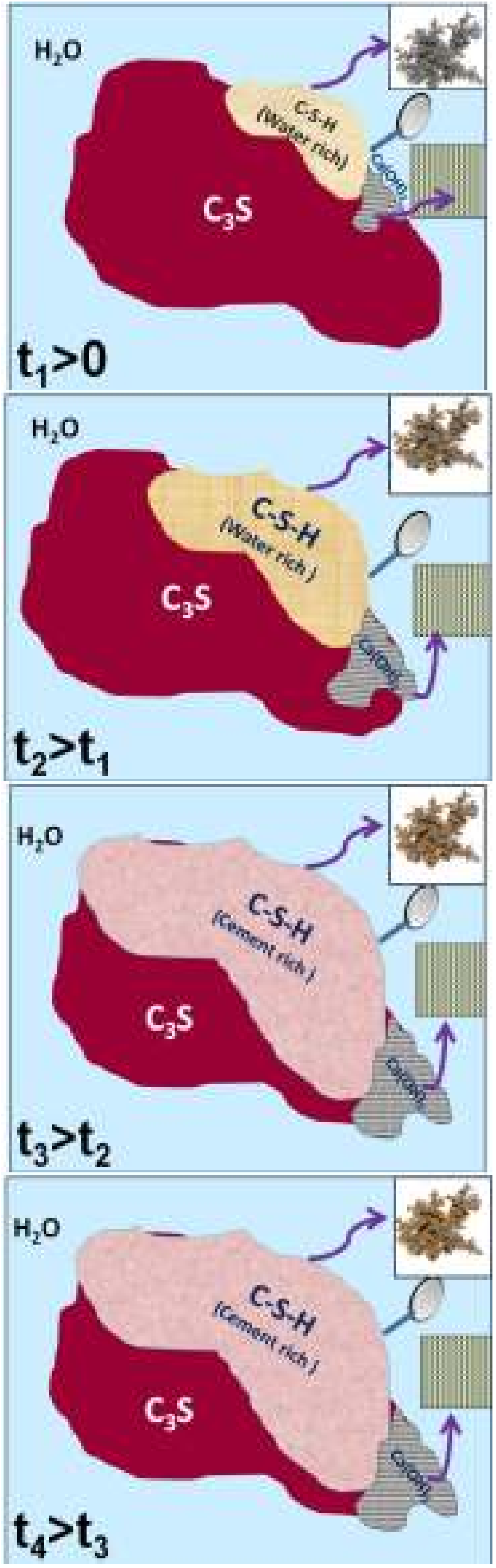

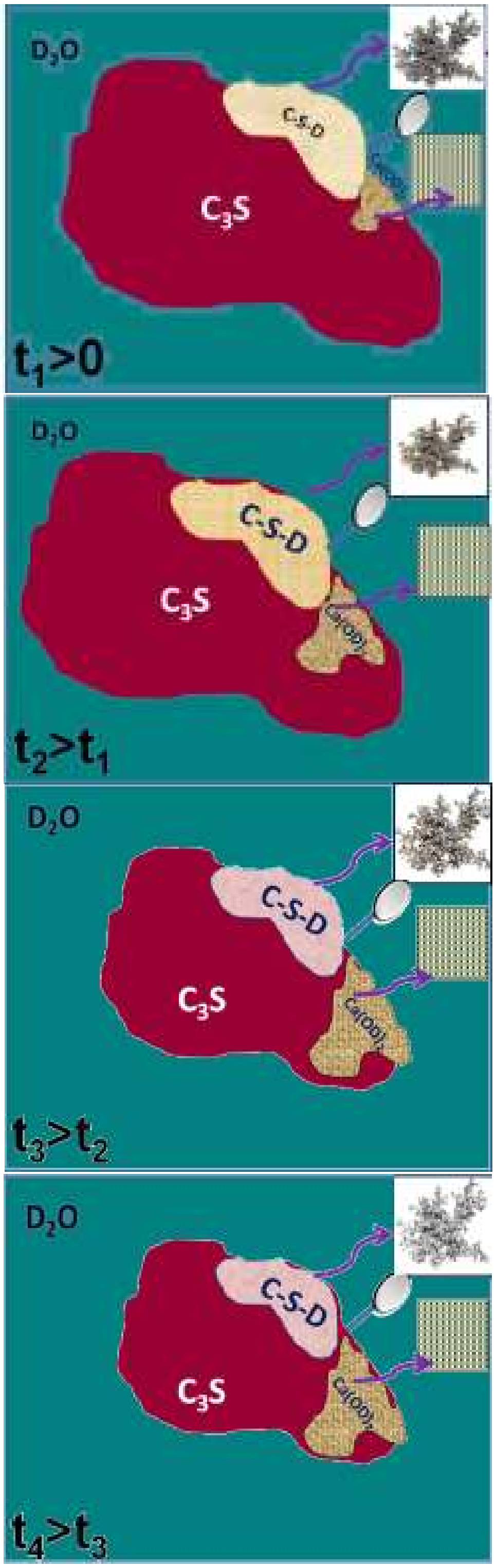

**Fig.3.** *$C_3S$, main constituent of Portland cement, (on top) and Temporal evolution of cement-water system during chemical reaction of $C_3S$ with light water (on left) and with heavy water (on right); Microstructures of products of the reactions are indicated at inset.*

Ordinary Portland cement is a composite material consisting of fine grains of tri-calcium silicate, $3CaO.SiO_2$ (Abbreviation $C_3S$; Approx. mass percentage range 60-80%) along with minor constituents like di-calcium silicate, tri-calcium aluminate, tetra-calcium iron aluminate etc. Hydration of $C_3S$, major component of Portland cement, by light water is described by following chemical reaction.

$$3CaO.SiO_2 + (3+y-x)H_2O \rightarrow (CaO)_x(SiO_2).(H_2O)_y + (3-x)Ca(OH)_2 \quad (1)$$

where $x$ is bounded by $0 \le x \le 3$ and y is bounded on one side i.e. $y \ge 0$. $x$=3 indicates hydration of $C_3S$ without formation of $Ca(OH)_2$. That $x$ is time dependent and functional form of *x(t)* is hydration medium ($H_2O$ or $D_2O$) dependent have been established[17] only recently. pH dependence of the kinetics of the reaction is a plausible reason for the time dependence of x(t). The product $(CaO)_x(SiO_2).(H_2O)_y$ (abbreviated as C-S-H, wherein hyphens indicate variable stoichiometry) is calcium silicate hydrate – a colloidal gel-like material.

In our model, we consider cement entirely made of $C_3S$, being major component of Portland cement, for simplicity. On mixing cement and light water, a complex series of hydration reactions[18] take place of which the main products are an amorphous calcium-silicate-hydrate (C-S-H) with gel-like structure and crystalline calcium hydroxide. The mesoscopic structure of C-S-H gel determines the desirable properties of hardened cement. The gel, having a non-Euclidean fractal morphology[8,19-21] constitutes about 60 -70 volume percentage of the fully hydrated cement paste which is a composite wherein unreacted cement powder and $Ca(OH)_2$ crystals are embedded as shown in Fig.3.

For a fractal system, a power law relation exists between various physical parameters over a wide length scale. The volume $V(r)$ of such an object varies as $r^{Df}$ where $D_f$ or fractal dimension < 3 over a wide range of length scale. But for a Euclidean object in three dimensions, $D_f = 3$. Further, fractal object that is generated following a particular mathematical rule is strictly self-similar and non-random in nature. However, for many natural and synthetic objects, the fractals are self-affine(i.e., have nonidentical scaling factors in different directions) and random in nature. A few of the computer models that have been proposed over last two decades to simulate such random fractals originated from the agglomeration of smaller particles are diffusion limited aggregation (DLA)[22] cluster-cluster aggregation[23,24], tunable dimension cluster-cluster aggregation[25,26], reaction limited cluster-cluster aggregation[27] etc. It is worth mentioning that the fractal dimension obtained from the SANS experiment just at the onset of hydration (both with $H_2O$ and $D_2O$) was found to be around 2.3. This prompted us to consider the DLA as a possible initial cluster in order to understand the dynamical evolution of the fractal structure.

Diffusion limited aggregation (DLA) is a process of random aggregation, the isotropic standard form of which is constructed by a purely diffusive algorithm. No hydrodynamics is involved in forming such an aggregate. However, an aggregate may be formed by combining diffusion with hydrodynamics in an appropriate manner like viscous finger growth. It is worth mentioning here that there exist many evolution models with different growth rules which yield DLA at large scale but, at a smaller length scale, structures could be different for different aggregation rules. However, many forms of diffusive growth with varying amounts of randomness exhibit DLA fractal morphology. Which growth rule is to be chosen is decided by the physical system that the aggregate is supposed to mimic. The structure of aggregate depends on how growth probability is distributed amongst different sites. What we need to determine is the growth probabilities of different sites for a given structure. This can be done for DLA and the way is to solve Laplace's equation for the potential when the cluster is considered to be made up of a conducting wire, kept at potential V. Then electric field at different tips gives the growth probability at these tips. This then is similar to solution of Laplace's equation in Darcy Law for viscous fingering aggregation. In Darcy flow, the governing equation is Darcy law, which states that the Laplacian of pressure is constant. The constant depends on the rate of flow and may be made negligible. Then the governing aggregation equation is the same as that governing DLA. Alternatively, the mean density profile of random DLA cluster may be related to the mean density profile of a viscous fingering pattern.

It is also assumed in DLA that the diffusing particles do not interact with one another in the sense that the motion of one particle is not affected by other particles. Each particle executes an independent random walk before sticking to the cluster growing around the seed or going outside the kill radius. If these diffusing particles move in a liquid, then the motion of one particle causes disturbances in the liquid which in turn affects the motion of other particles. If this process, known as hydrodynamic effect of interaction, is important, then the motions of different particles cannot be thought of as being independent of one another - hydrodynamic interactions induce correlations in motions of different particles.

But in the present problem, cement particles in contact with water molecules undergo hydration, and they do not have to diffuse through water over long distances. Hence, ignoring hydrodynamic effect is justified. Further, the probability of sticking at different points in DLA is averaged over many evolutions with the same starting cluster, and the growth probability satisfies Laplace’s equation with certain boundary conditions. The interaction of diffusing particles with one another or with molecules of the liquid medium is not required to determine growth probability of a DLA cluster at different points along its perimeter as it is akin to the velocity of growth in the cluster of same geometry in viscous fingering.

In the present work, simulation of a DLA cluster starts by keeping one particle at the origin of a lattice (this particle will be termed as seed) and allowing another particle, released from some release radius ($r_0$), to execute random walk until it either touches the seed or goes out of some kill radius ($r_k$). A number of particles are thus released and some of them are finally aggregated around the seed to form a self repetitive structure. For faster convergence of the simulation, some critical radius ($r_c$) is defined, beyond which particles take longer steps while after stepping into the radius they start taking shorter steps for walk. $r_o$ , $r_k$ and $r_c$ can be adjusted intuitively to speed up the process. After obtaining a realizable DLA cluster both in 2D and 3D (Fig.4.(a) and (b)), hydration reaction with light water was first simulated.

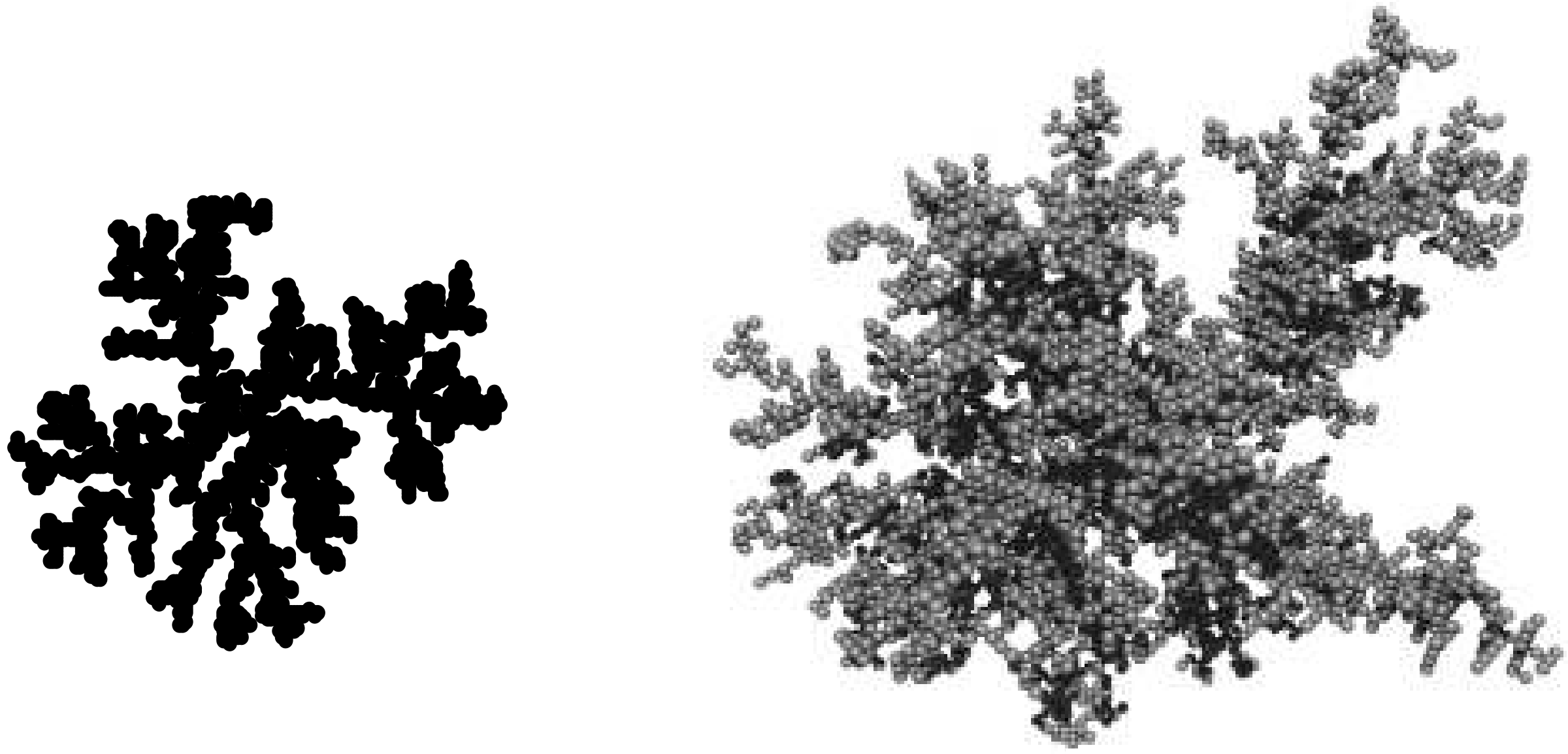

*(a)* *(b)*

**Fig.4.** *Initial DLA clusters in (a) two and (b) three dimensions.*

As soon as the hydration process starts, gel begins to be formed and as time passes this gel fills the available space in between the cluster to make a consolidated structure. However, it is understood that the growth of the cluster is constrained because of the similar response from adjacent clusters.

For the present calculation, it was considered that the DLA structure is surrounded by light water molecules. Though light water molecules were not represented explicitly, their reaction with $C_3S$ was manifested in following way. Firstly, one site was selected randomly from all the possible sites and then the cement particle occupying that site was replaced with product particles (for simplicity will be called later as gel particle) composed of gel and hydroxide. This is how one Monte Carlo (MC) step was completed. After repeating the same mechanism at various other sites, a layer of products was formed on the cluster surface at the cost of cement particles. This layer restricts further reaction of water with cement and hence, probability of hydration reaction gets reduced. Fractal dimension ($D_f$) and characteristic length ($L$), giving a true flavor of system characteristic, were evaluated with a regular interval of MC step. Mathematically, fractal dimension ($D_f$) is defined as following:

$$D_f = \frac{log_{10}\left(N\left(\varepsilon\right)\right)}{log_{10}\left(\frac{1}{\varepsilon}\right)} \tag{2}$$

$N\left(\varepsilon\right)$ is the number of self similar structures of linear size $\varepsilon$ required to cover the entire fractal object. Numerically $N\left(\varepsilon\right)$ was calculated by box counting method. Characteristic length was calculated from radius of gyration ($R_g$) as defined below for the entire cluster with respect to an arbitrary origin

$$R_g^2 = \frac{\int_V \rho(r) r^2 d^3 r}{\int_V \rho(r) d^3 r} \tag{3}$$

It is to be noted that in earlier experiments[2], radius of gyration was estimated from SANS data using scattering length density (SLD) as the weighting function of the system. It can be shown that the SLD of tri-calcium silicate ($C_3S$) (~$3.94\times10^{14}$ $m^{-2}$) is greater than that of C-S-H gel (~$2.29\times10^{14}$ $m^{-2}$) and lesser than that of C-S-D gel (~$4.28\times10^{14}$ $m^{-2}$)[17] which are taken into consideration during the present simulation. In this model, SLD of gel particle was scaled with respect to SLD of cement particle.

In case of hydration with heavy water, similar kind of chemical reaction (i.e. Eq.1) as in light water case is responsible where C-S-D gel and $Ca(OD)_2$ are formed and it was incorporated into the simulation in identical manner. But due to stronger hydrogen bond – a chemical effect indeed, these newly formed C-S-D gel particles coalesce and a new kind of consolidated gel particle is formed. This was simulated by replacing a randomly picked up cluster of product particles with a single particle, conserving the mass.

At the beginning of hydration, gel particles are less in number and therefore probability that the gel particles coalesce into one is low. With time this probability goes up as total number of gel particles formed in the hydration reaction increases. But after sometime the probability comes down also as very few gel particles will be left behind in the cluster to shrink.

Evolution of $D_f$ and $(L(t))^2$ with MC step for light water and heavy water are shown in Fig.5, Fig.6 and Fig.7.

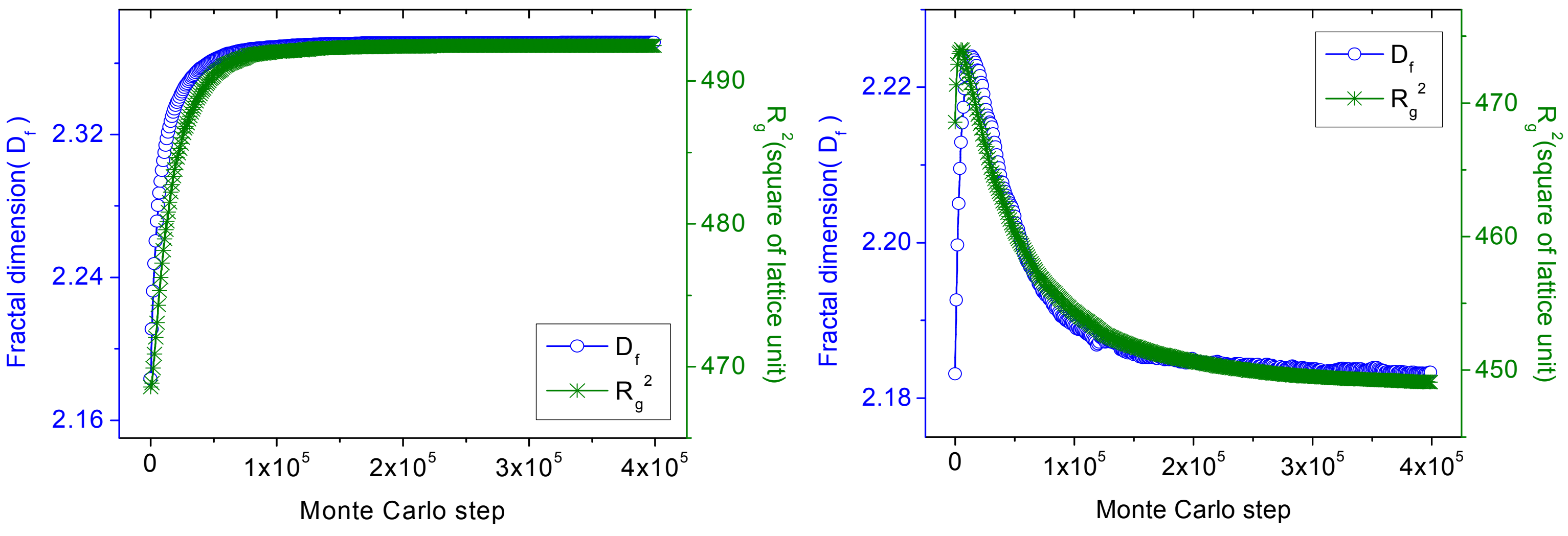

*(a)* *(b)*

**Fig.5.** *(a) Variations of fractal dimension and characteristic length square with Monte Carlo step in light water hydration case. (b) Variations of fractal dimension and square of characteristic length with Monte Carlo step in heavy water hydration case. In (a) and (b) both, sticking probability was taken as unity.*

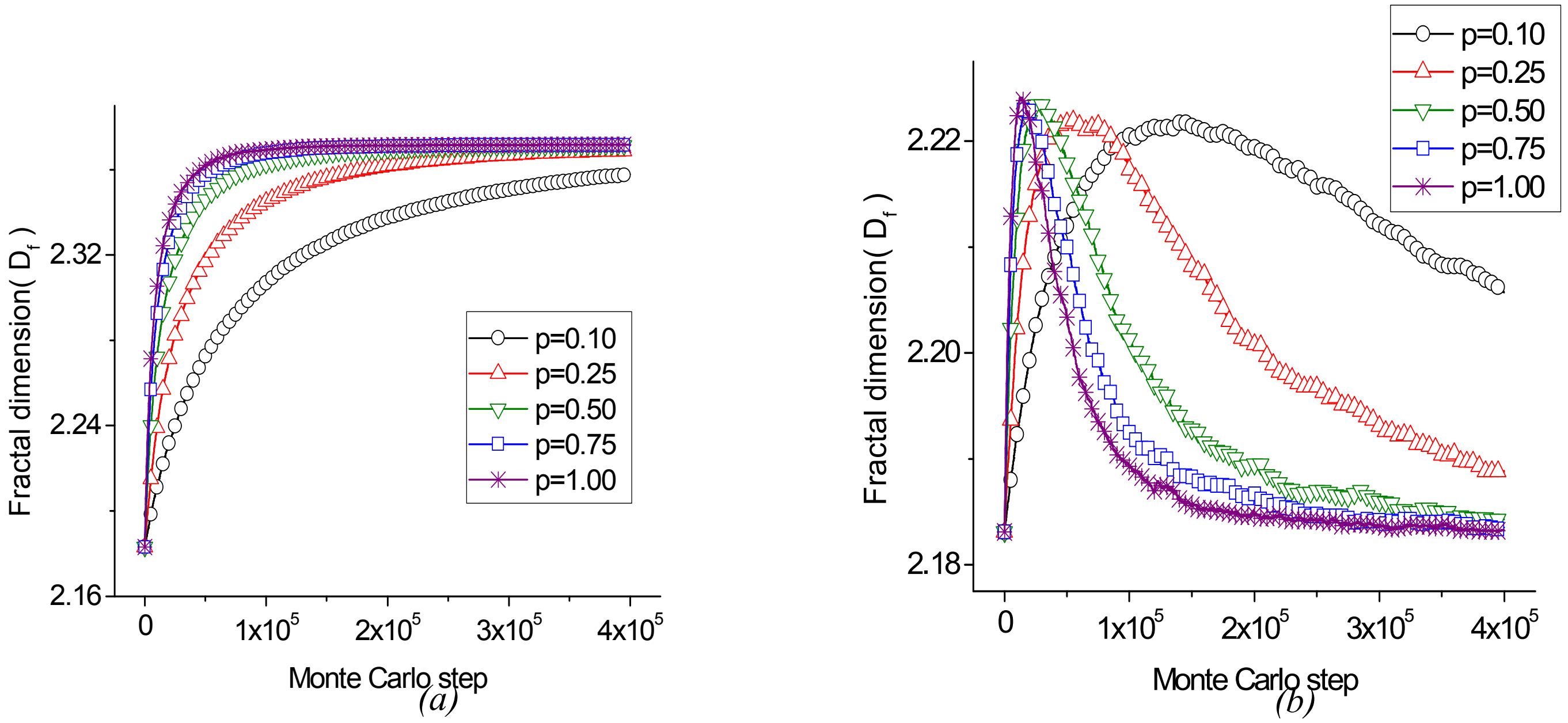

*(a)* *(b)*

**Fig.6.** *(a) Variations of fractal dimension with Monte Carlo step for different sticking probabilities in light water hydration case. (b) Variations of fractal dimension with Monte Carlo step for different sticking probabilities in heavy water hydration case.*

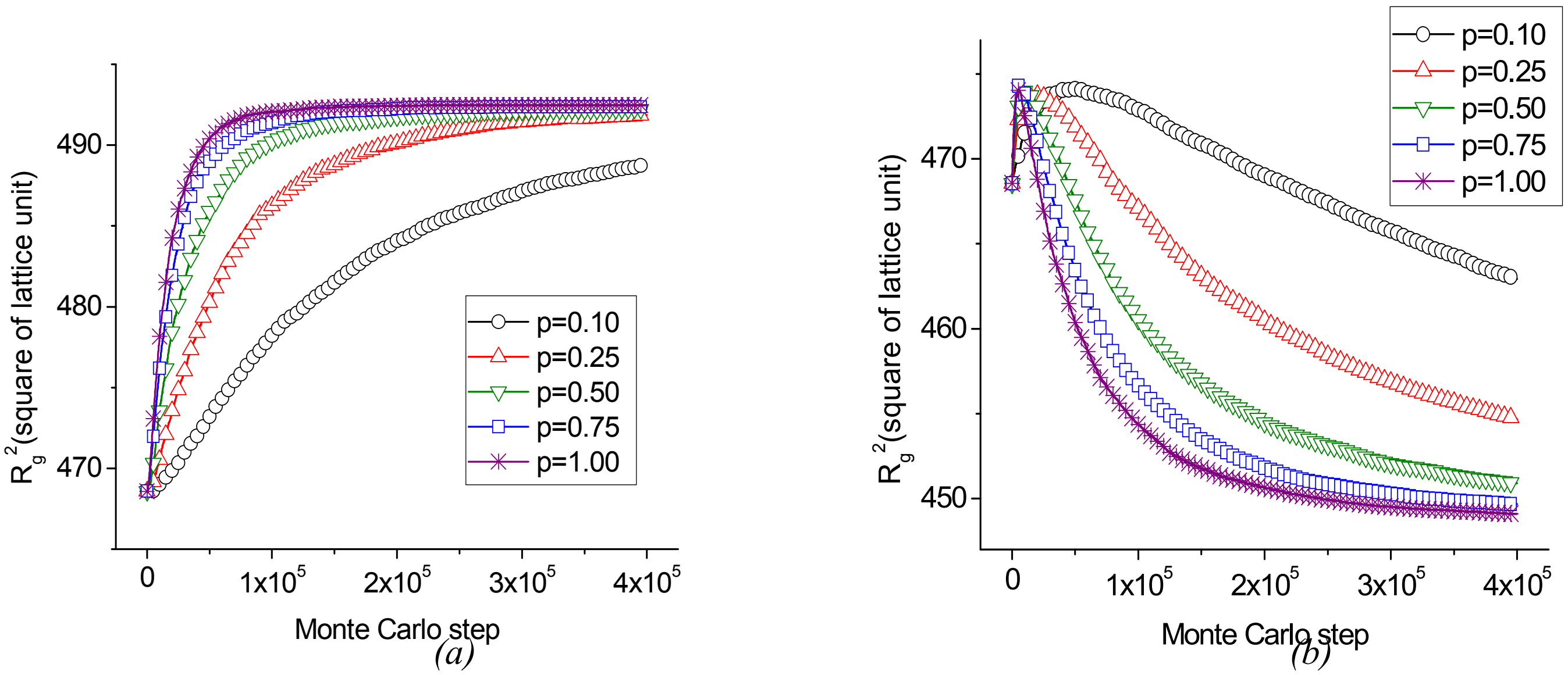

**Fig.7.** *(a) Variations of square of characteristic length with Monte Carlo step for different sticking probabilities in light water hydration case. (b) Variations of square of characteristic length with Monte Carlo step for different sticking probabilities in heavy water hydration.*

**3. Results and discussion:**

Fig.4 depicts the initial DLA clusters, made by the random walk approach, in two as well as in three dimension. As in the figure, the cluster is made of ~3200 cement particles for 2D case and ~5800 particles for 3D case. 2D cluster is introduced here just to show the morphological difference between 2D and 3D clusters. In present study, hydration reactions with light water and heavy water were simulated for 3D cluster only as described in the previous section. The probability for a gel particle to be a part of existing cluster is termed as sticking probability *p*. For the present simulation, various values namely, 0.10, 0.25, 0.50, 0.75 and 1.00, of *p* have been considered. In case of hydration with light water, gel particles try to fill the available space in and around the cluster. Therefore, an initial non-Euclidean morphology of the cluster tends towards a consolidated Euclidean morphology with the progress in hydration process. From Fig.5.(a), it is evident that the value of $D_f$ of the cluster remains ~2.18 just at the onset of hydration. Thereafter, $D_f$ increases with time i.e. the cluster tends towards an Euclidean morphology. At around $5\times10^4$ MC steps, it reaches a plateau. This happens because the gel particles form a protective layer on the surface of cluster and hence, hydration reaction gets seized unless and until there is a rupture of the protective layer. $R_g^2$ of the cluster also increases with time and finally reaches a plateau almost at the same time as $D_f$ (Fig.5.(a)). Since SLD of C-S-H gel is less than that of $C_3S$, therefore the cluster expands volumetrically as soon as the gel particles begin to form on initial cluster. But the rate of expansion gradually gets reduced due to the protective layer which hinders further gelation. Since the present computational work deals with a single cluster, therefore it is difficult to realize the same percentage rise in $R_g^2$ as observed experimentally. At the onset of hydration, neutrons, used as a probe in the experiment, are able to see all the clusters isolated. Due to incessant gelation, clusters grow with time and gradually they are joined together to form a bigger cluster. As a consequence, neutrons see only this bigger cluster which is responsible for increase of $R_g^2$ with time in experiment.

In case of hydration with heavy water, unlike the case of hydration with light water, the gel particles are formed first and then shrink due to stronger hydrogen bond. Initially, the rate of shrinkage is dominated by the rate at which gel particles form. So at the onset of hydration, both $R_g^2$ and $D_f$ increase, reach a peak, after which shrinkage rate starts to surpass the formation rate and hence the curve falls down i.e., the cluster again goes towards non Euclidean morphology with increase of deviation from the corresponding Euclidean morphology. From Fig.5.(b), it is seen that $D_f$ reaches the peak at about $1.4\times10^4$ MC steps and after that gradually decreases with time. Similar trend is observed for $R_g^2$ also (Fig.5.(b)). Stronger hydrogen bonds involving

Deuterium atoms lead to consolidation of gel at the later stage for hydration of cement with heavy water. This leads to $R_g^2$ decreasing with time. But its percentage rise and fall do not corroborate exactly with experimental observation due to the reason mentioned earlier.

One outcome of the present simulation is the effect of sticking probability $p$ on the time at which $D_f$ or $R_g^2$ reaches plateau (in case of $H_2O$) or attains a maximum value (in case of $D_2O$). Smaller sticking probability implies less affinity of sticking of the gel particles with the cement particles. Hence, the cluster evolves with a slower kinetics and respective curves reach their plateau or peak at a later time. In Figs 6.(a) and 7.(a), $D_f$ and $R_g^2$, respectively, reach their plateau for $p$=1.00, 0.75 and 0.50, whereas the plateau is not reached for the cases with $p$=0.25 and $p$=0.10. In Fig.6.(b) and Fig.7.(b), the peaks are shifted towards higher MC steps with the decrease in $p$ value.

As the theme of the present study was motivated from earlier results[1,2,9,16] of small angle scattering function, the evolution of scattering function for the cluster has been studied here. studied here. The scattering function $S(q,t)$ is expressed as

$$S(q,t) = CP(q)Q(q,t) \tag{4}$$

where $C$ is a scale factor and independent of $q$, for simplicity $P(q)$ was taken as a form factor of a sphere with radius $\tilde{r}_0$

$$P(q) = \frac{[\sin(q\tilde{r}_0) - q\tilde{r}_0 cos(q\tilde{r}_0)]^2}{(q\tilde{r}_0)^6} \tag{5}$$

and $Q(q,t)$ was taken as that for a mass fractal[12, 28]

$$Q(q,t) = 1 + \frac{1}{(q\tilde{r}_0)^{D_f(t)}} \frac{D_f(t)\Gamma(D_f(t)-1)}{[1+\frac{1}{q^2\zeta^2}]^{\frac{(D_f(t)-1)}{2}}} \sin[(D_f(t)-1)\tan^{-1}(q\zeta)] \tag{6}$$

where $\zeta$ is the upper cutoff of the fractal and has been taken same as the characteristic length scale. The evolution of scattering functions for the $H_2O$ and the $D_2O$ cases are depicted in the Figs.8.(a) and (b), respectively. The insets show the variation of the scattering intensity at $q$~0. It is interesting to note from the figures that the scattering profiles evolve differently with time for $H_2O$ and $D_2O$ cases. The evolutions resemble the trends as observed in the earlier scattering experiments.

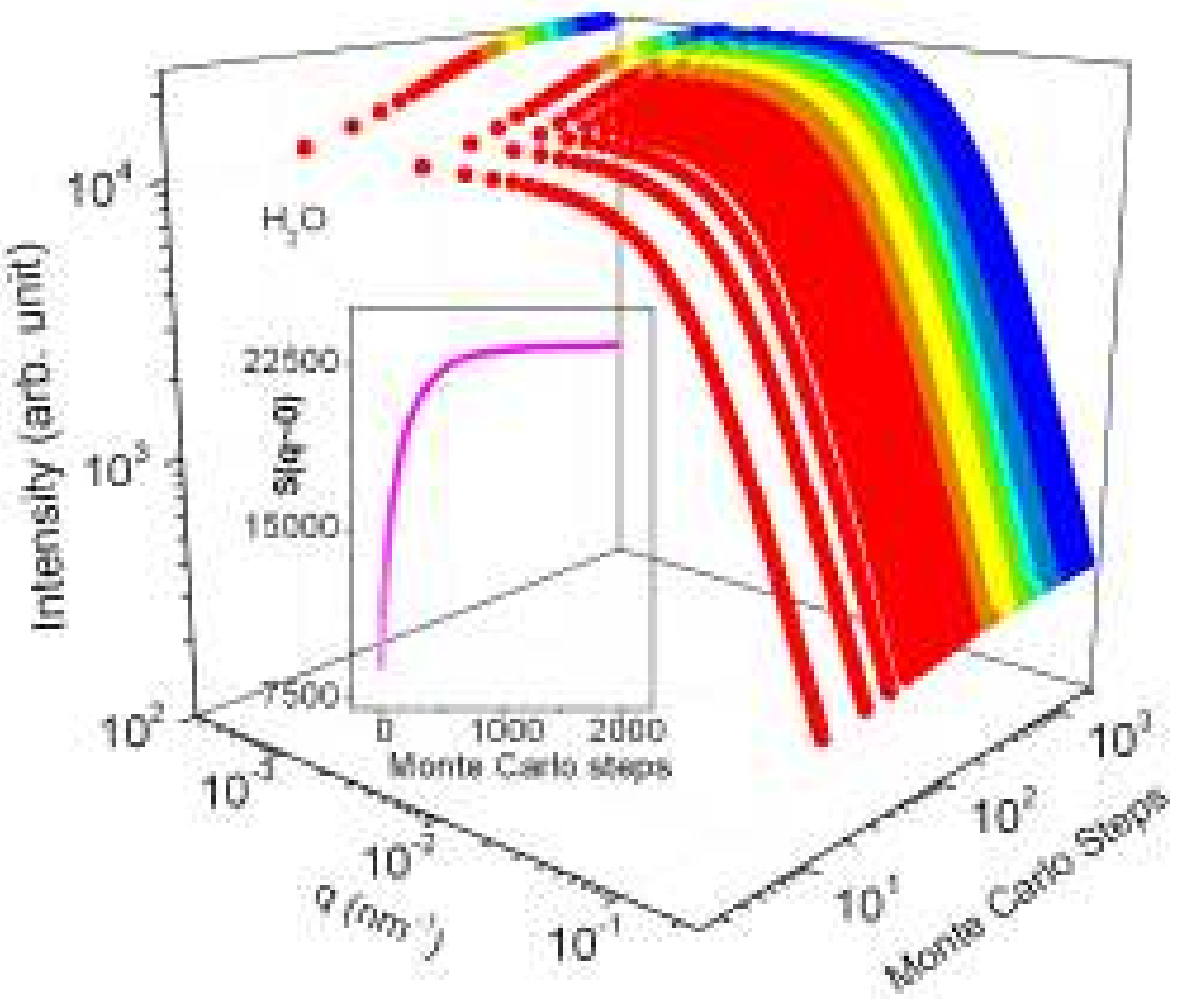

*(a)*

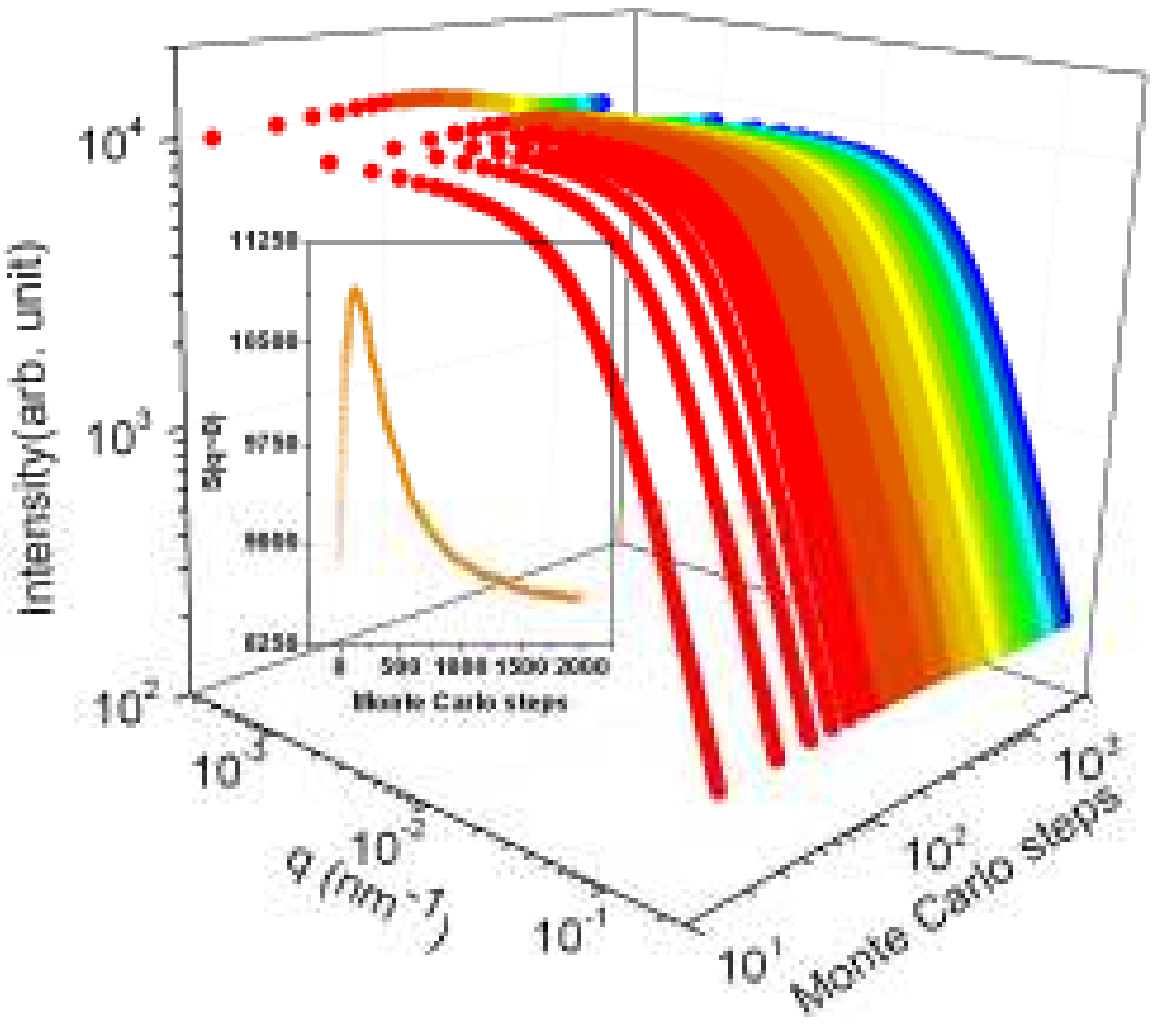

*(b)*

**Fig.8.** *(a) Evolutions of the simulated scattering profile with Monte Carlo steps at different q values for light water hydration case. (In inset) Evolution of simulated scattering intensity with Monte Carlo steps at q~0 for light water hydration case. (b) Evolutions of the simulated scattering profile with Monte Carlo steps at different q values for heavy water hydration case. (In inset) Evolution of simulated scattering intensity with Monte Carlo steps at q ~ 0 for heavy water hydration case.*

Fig.9 shows the morphological difference between clusters while hydrated with light water and heavy water. At $10^3$ MC steps, there is no significant difference between them as both the clusters grow with time due to gel formation. But in subsequent MC steps, since C-S-D gel starts to shrink whereas C-S-H gel continues to be formed without any contraction of matrix, the resultant clusters become different. It is evident from the figure that the light water hydrated cluster grows with time while the heavy water hydrated cluster, in spite of a little sign of growth at the onset of hydration, shrinks finally in size.

At this juncture, let us try to understand the simulation results on the basis of a proposed growth model associating the rate equation. Let $N_{ci}(t)$ be the number of cement particles situated inside the cluster, $N_{cs}(t)$ be the number of cement particles at the boundary surface of cluster and $N_g(t)$ be the number of gel particles. Due to the presence of protective layer at the boundary, $N_{ci}(t)$ does not change with time and hence, $N_{ci}(t)$ is written as $N_{ci}$.

For the present model calculation, it is assumed $(3+y-x)=1$ in chemical reaction (i.e. Eq.1) of hydration of cement, a plausible case indeed. Therefore, temporal variation of $N_{cs}(t)$ is dictated by following rate equation:

$$\frac{dN_{cs}(t)}{dt} = -K_{c\to g} N_{cs}(t) N_w(t) \tag{7}$$

where $K_{c\to g}$ is the rate constant of above reaction and $N_w(t)$ represents the number of light water molecules. The assumption $(3+y-x)=1$ implies $N_w(t) - N_{cs}(t) = c$, where $c$ is a time-independent constant which depends on initial concentrations of reactant molecules. Now, replacing $N_w(t)$ by $(N_{cs}(t) + c)$ in Eq.(7) and solving the differential equation, we have

$$N_{cs}(t) = \frac{c}{\left(1 + \frac{c}{N_{cs}(0)}\right)e^{cK_{c \to g}t} - 1} \tag{8}$$

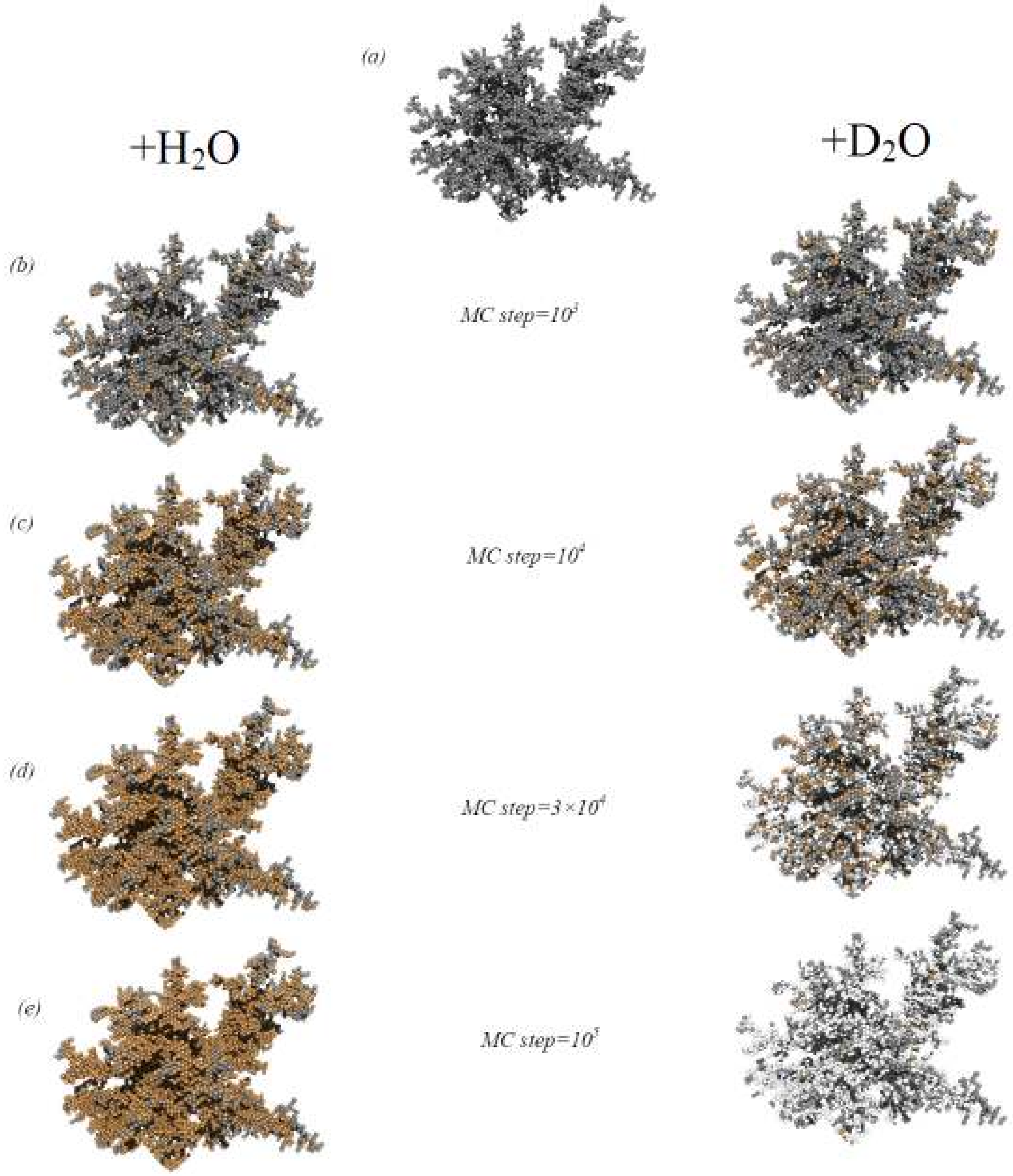

**Fig.9.** *(a) Initial DLA cluster. (b)-(e) Evolution of DLA cluster while hydrated with light water(on left) and heavy water(on right) at different MC step indicated in the figure.*

The present simulation assumes that gel particles are formed with a higher rate compared to the depletion rate of cement particles in order to account the volumetric expansion of cluster. Since $N_g(0)=0$, therefore

$$N_g(t) = \tilde{\eta}\left(N_{cs}(0) - N_{cs}(t)\right) \tag{9}$$

where $\tilde{\eta}$ is the ratio of formation rate of gel particles and depletion rate of cement particles.

Total number of cement and gel particles at time $t$ becomes

$$N(t) = N_{ci} + \tilde{\eta}N_{cs}(0) + \frac{c(1-\tilde{\eta})}{\left(1 + \frac{c}{N_{cs}(0)}\right)e^{cK_{c \to g}t} - 1} \tag{10}$$

Therefore, fractal dimension($D_f$) can be written as

$$\begin{aligned} D_f(t) &= \frac{log_{10}\left(N(t,\varepsilon)\right)}{log_{10}\left(\frac{1}{\varepsilon}\right)} \\ &= L\,log_{10}\left(N_{ci} + \tilde{\eta}N_{cs}(0) + \frac{c(1-\tilde{\eta})}{\left(1 + \frac{c}{N_{cs}(0)}\right)e^{cK_{c \to g}t} - 1}\right) \end{aligned} \tag{11}$$

where, $L = \left[log_{10}\left(\frac{1}{\varepsilon}\right)\right]^{-1}$.

Fig.5.(a) can be fitted with Eq.(11) (Fig.10). Hence, the above equation is proposed as a fitting function for the curve, $D_f$ vs. MC step in case of light water.

It is evident from Fig.10 that the variation of $D_f$ with Monte Carlo step obtained from simulation, so the corresponding experimental observations also, corroborates well with the temporal variation of $D_f$ predicted by the growth model for the case of hydration of cement with light water. In this context, the relation between MC step and real time of hydration is addressed in Fig. 11. Identical change in $D_f$ is first identified in both experiment and simulation and then the corresponding change in time and MC steps are worked out respectively. This equivalence helps us to relate the computer model with the real hydration process and hence to confirm that the model is realistic.

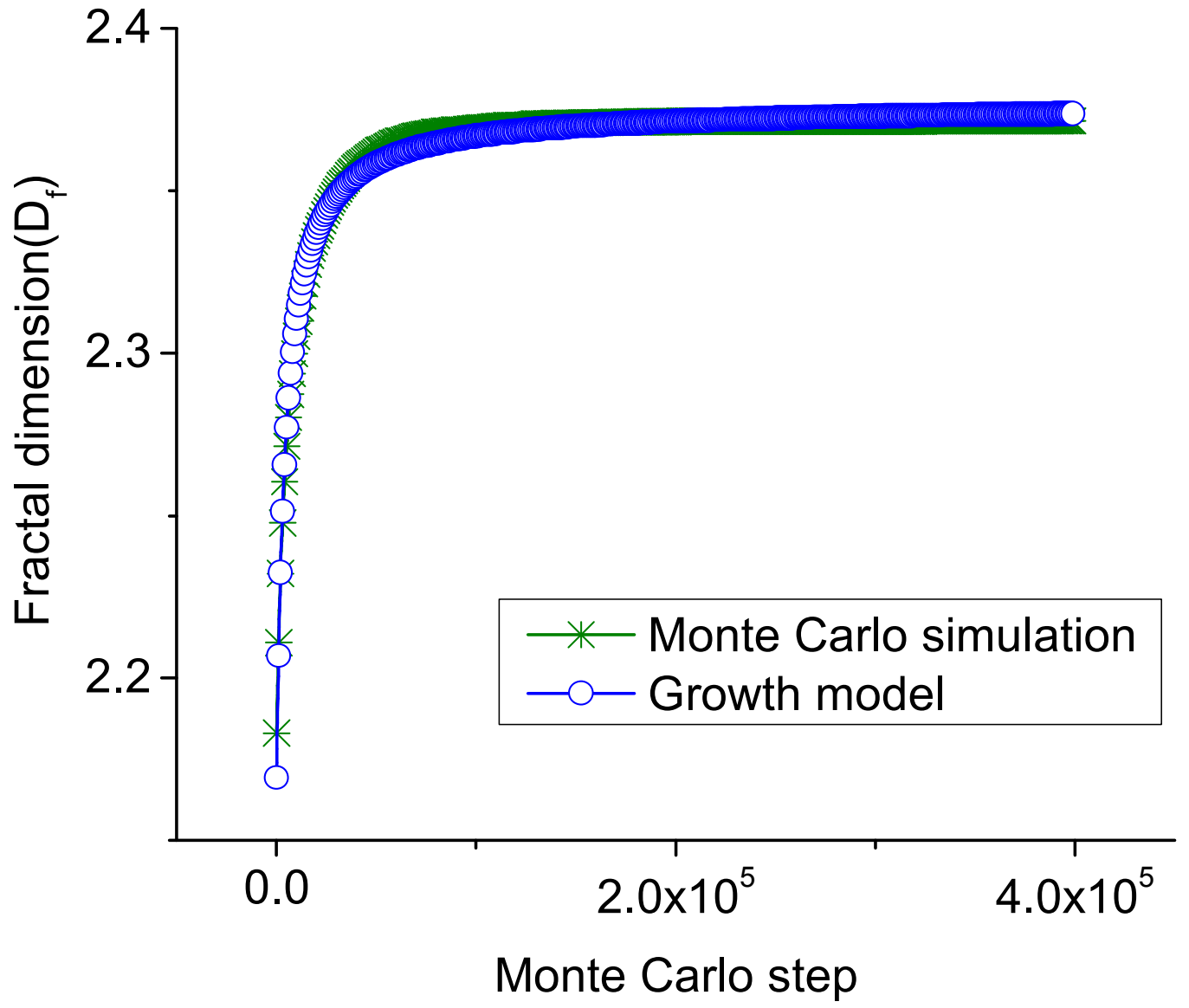

**Fig.10.** *Fitting of $D_f$ vs. number of MC steps for light water hydration of cement as in Fig.4.(a) with Eq.(10) in the proposed growth model for L~0.26, $N_{ci}$~$6.72\times10^2$, $\tilde{\eta}$~2.45, $N_{cs}(0)$~$3.83\times10^3$, c~$1.00\times10^{-5}$, $K_{c\rightarrow g}$~$3.81\times10^{-8}$ min.$^{-1}$. This is not a unique set of values. It varies depending on the guessed values provided before the fitting starts, which implies that the same temporal variation of $D_f$ is possible for clusters of different morphology.*

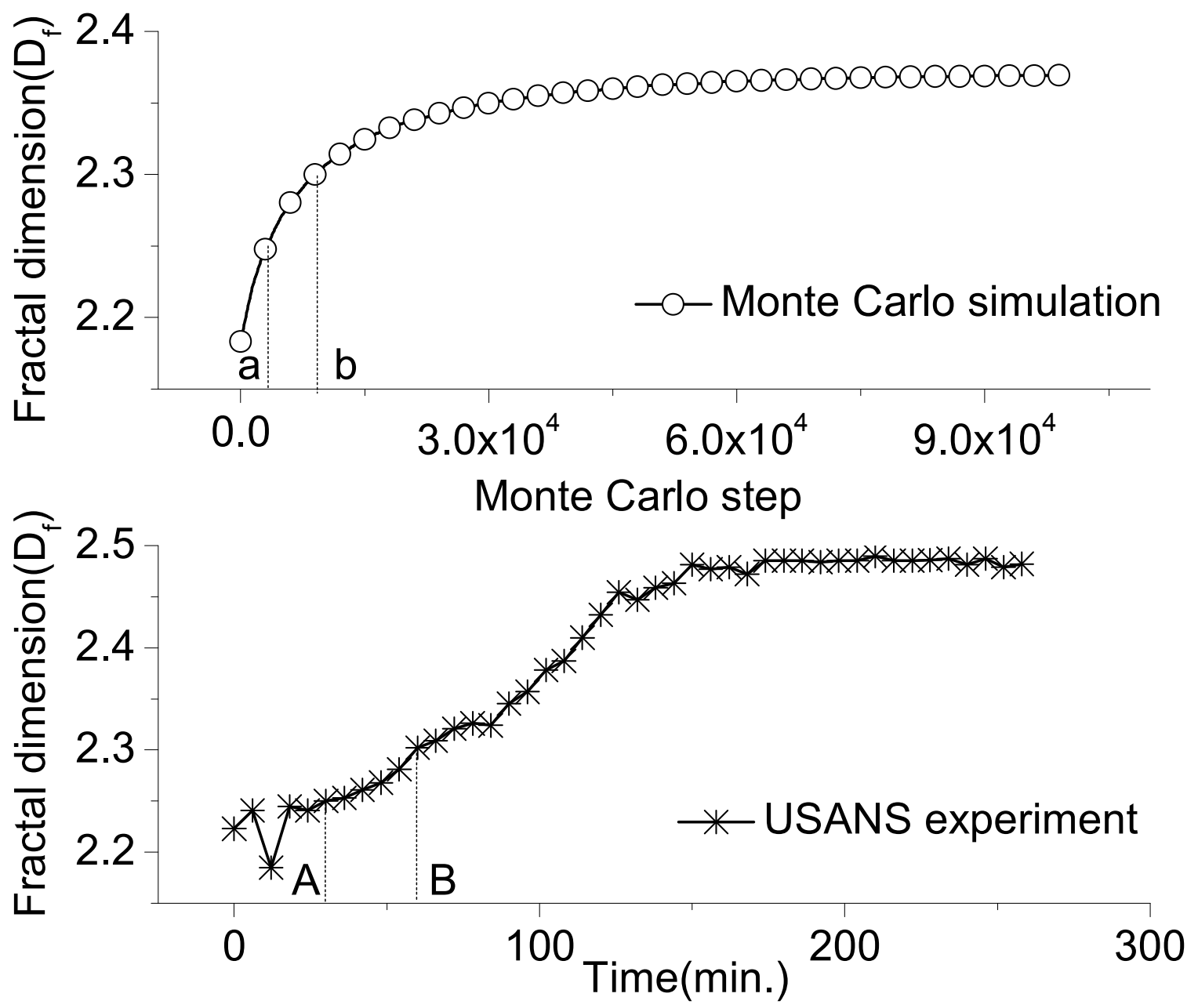

**Fig.11.** *Variation of $D_f$ with number of MC steps and real time of hydration in the case of light water hydration. In the dotted sections, $D_f$ changes from 2.25 to 2.3 for both cases and the corresponding change in number of MC steps is (b-a)=5800 MC steps and in real time is (B-A)=30 min. (b-a)/(B-A)~193 MC steps/min i.e. the change that 193 MC steps bring about into the simulated system, occurs in 1 min. in the real system. But this factor varies depending on the state of the system in different time regimes.*

In case of hydration of cement by heavy water, C-S-D gel is not only formed, but also shrinks. Let us assume that $\xi$ number of gel particles coalesce to form a single consolidated gel particle. Hence, this two step reaction can be simply written as $c \rightarrow \tilde{\eta}\, g$ in first step as in the earlier case of hydration and $\xi\, g \rightarrow G$ in second step where, *c,g and G* represent cement, C-S-D gel and consolidated gel, respectively. Therefore,

$$\frac{dN_g(t)}{dt} = -\tilde{\eta}\frac{dN_{cs}(t)}{dt} - K_{g\to G}N_g^{\xi}(t) \tag{12}$$

where $K_{g\to G}$ is rate constant for the second step of reaction.

Now, the formation rate of consolidated gel particle is given by

$$\frac{dN_G(t)}{dt} = \frac{1}{\xi}K_{g\to G}N_g^{\xi}(t) \tag{13}$$

where $N_G(t)$ is the number of new gel particles and $1/\xi$ is the ratio of formation rate of new gel particle and the depletion rate of precursor. The temporal variation of $N_g(t)$ and $N_G(t)$ are shown in Fig. 12.

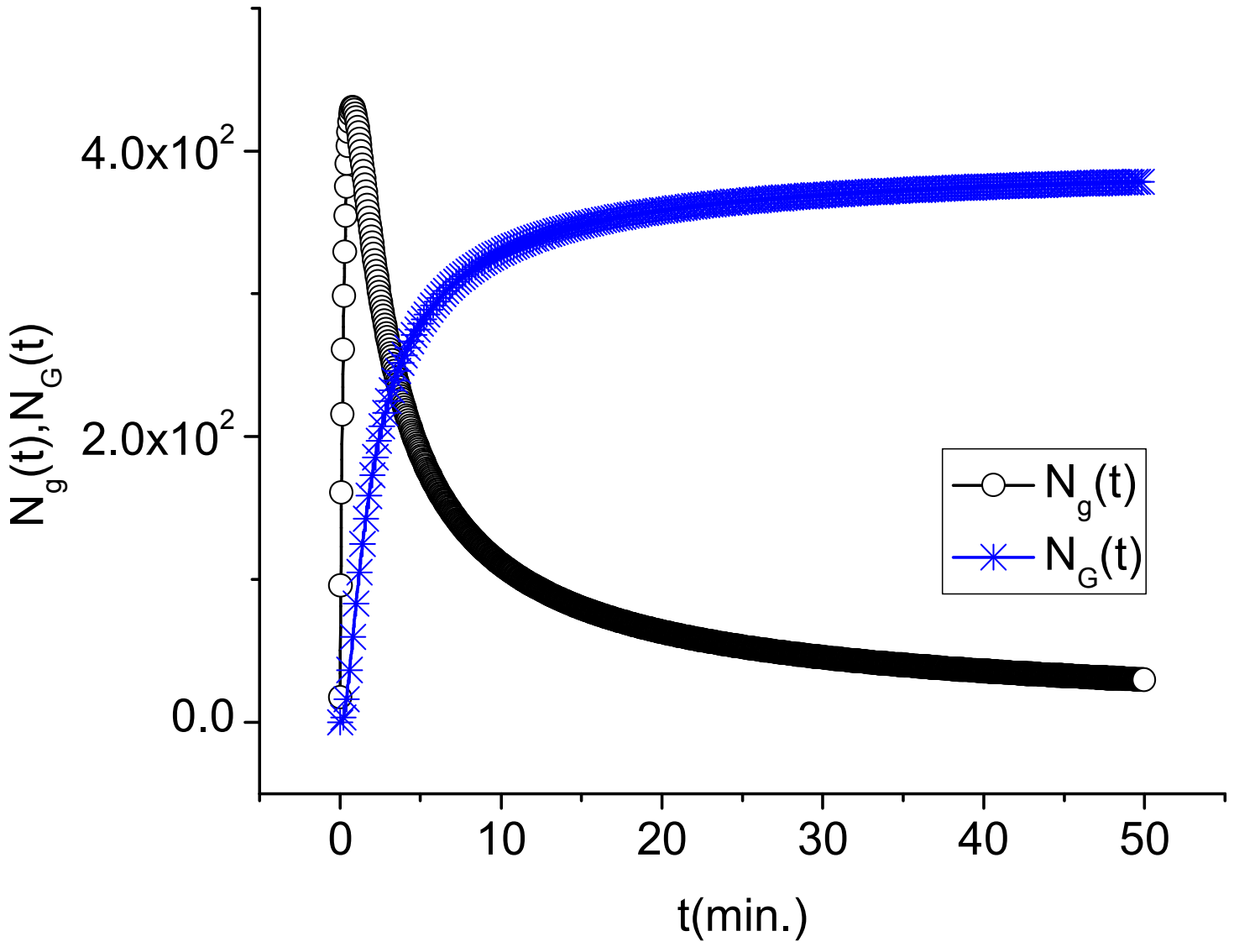

**Fig.12.** *$N_g(t)$ vs. t and $N_G(t)$ vs. t as proposed in growth model with* $\tilde{\eta}$ *~ 2.45, $N_{cs}(0)$~$3.83\times10^3$, c~$1.00\times10^{-5}$, $K_{c\to g}$ ~$4.85\times10^{-3}$min.$^{-1}$, $K_{g\to G}$~$3.35\times10^{-4}$min.$^{-1}$, $\xi$~2.25. This is not a unique set of values for the reason already mentioned in Fig.10.*

In this case, fractal dimension($D_f$) is expressed as

$$D_f(t) = \frac{log_{10}\left(N(t,\varepsilon)\right)}{log_{10}\left(\frac{1}{\varepsilon}\right)}$$

$$= L\,log_{10}\left( N_{ci} + \frac{c}{\left(1+\frac{c}{N_{cs}(0)}\right)e^{cK_{c\to g}t} - 1} + N_g(t) + N_G(t) \right) \qquad (14)$$

where, $L = \left[ log_{10}\left(\frac{1}{\varepsilon}\right) \right]^{-1}$.

Eq.(14) reproduces the same kind of variation of $D_f$ with time(in experiment) as well as with MC steps(in simulation) observed in heavy water hydration of cement(Fig.13).

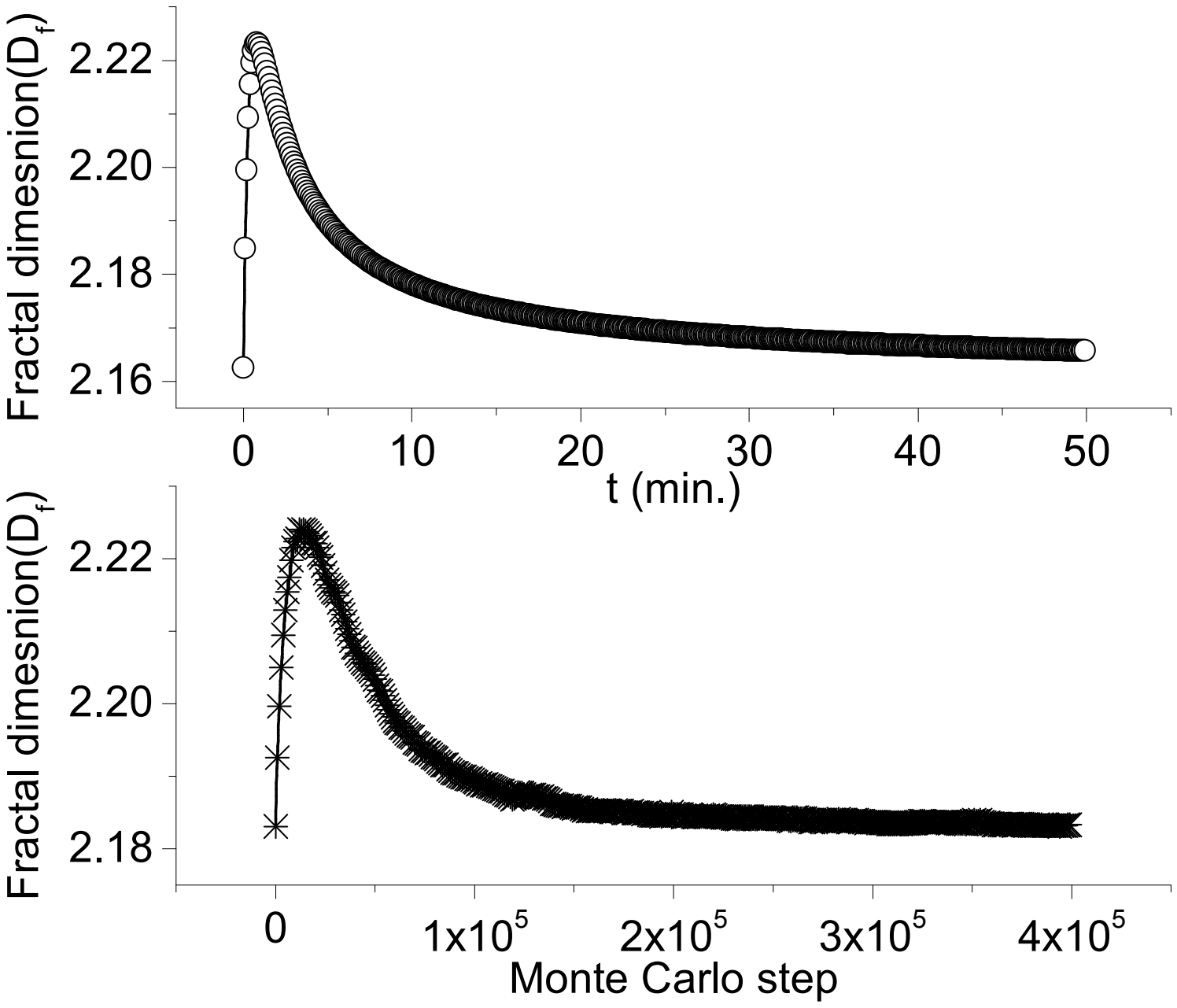

**Fig.13.** *$D_f$ vs. t as proposed in growth model with $L$~0.31, $N_{ci}$~$7.20\times10^2$, $\tilde{\eta}$~2.45, $N_{cs}(0)$~$3.83\times10^3$, c~$1.00\times10^{-5}$, $K_{c\to g}$~$4.85\times10^{-3}min^{-1}$, $K_{g\to G}$~ $3.35\times10^{-4}min^{-1}$, $\xi$ ~ 2.25*

*(top). $D_f$ vs. number of MC steps for heavy water hydration of cement by computer model (bottom). This is not a unique set of values for the reason already mentioned in Fig.10.*

It is evident from Fig.13 that the variation of $D_f$ with Monte Carlo steps obtained from simulation, so the corresponding experimental observations also, corroborates well with the temporal variation of $D_f$ predicted by the growth model for the case of hydration of cement with heavy water.

**4. Conclusions:**

In accordance with present Monte Carlo based computer model, space filling mechanism of C-S-H gel is the key to analyze the temporal variation of fractal dimension and characteristic length in case of light water hydration whereas for the hydration with heavy water, both filling and shrinkage of C-S-D gel are responsible for the contrasting temporal behavior of above mentioned physical quantities. Radius of gyration, which involves scattering length density of gel, was considered to be the characteristic length in the simulation. A growth model was proposed to explain only the temporal behavior of fractal dimension for both light water and heavy water hydration cases and it corroborates well with the experimental observations and Monte Carlo simulation results. Equations coming out from the model (i.e. Eq.11 and 14) may look complicated, but suitable mathematical adjustments (like merging two/three parameters into one etc.) can make them simpler. But any such mathematical simplification was not done since the focus was to find out the best fitted values of each and every parameter involved in the equations so that, if any measurement of those parameters is done in future, the measurement results can be linked with these best fitted values. Hence, the present calculations demand further experiments in future. Moreover, an extension of the present theoretical model for the evolution of radius of gyration during light and heavy water hydration of cement will be considered in near future. It is important to mention that the entire analysis is applicable for hydration observed in short time scale. Its long time behavior still remains unknown and hence it demands more detailed scattering investigations.

## 5. Acknowledgement:

The authors are grateful to the reviewers for their kind suggestions to improve the quality of presentation. We also acknowledge the discussion one (SM) of us had with Prof. D. Dhar in regard to the computer modeling.

**Figure captions:**

**Fig.1.** *(a) Temporal Evolution of $\eta$ for both light water and heavy water hydrating silicates as observed from neutron scattering measurements*[1]*. Only a typical representative figure has been considered. (b) Temporal Evolution of* $(L(t))^2$ *for both light water and heavy water hydrating silicates as observed from neutron scattering measurements*[1]*. Only a typical representative figure has been considered. For light water hydrating silicates, both $\eta$ and* $(L(t))^2$*increase with time and reach a plateau whereas for heavy water hydrating silicates, they first increase, then reach a peak and finally decrease as time approaches.*

**Fig.2.** *Temporal Evolution of* $S(q_{min}(t),t)$ *for light water and heavy water*[1] *hydrating silicates where* $S(q,t)$ *represents scattering function and* $q_{min}$ *is the lowest attained q value for a particular measurement. It resembles the nature of temporal evolution of $\eta$ and* $(L(t))^2$ *for light and heavy water hydration(Fig.1). We have considered only a typical representative figure here.*

**Fig.3.** *$C_3S$, main constituent of Portland cement, (on top) and Temporal evolution of cement-water system during chemical reaction of $C_3S$ with light water (on left) and with heavy water (on right); Microstructures of products of the reactions are indicated at inset.*

**Fig.4.** *Initial DLA clusters in (a) two and (b) three dimensions.*

**Fig.5.** *(a) Variations of fractal dimension and characteristic length square with Monte Carlo step in light water hydration case. (b) Variations of fractal dimension and square of*

*characteristic length with Monte Carlo step in heavy water hydration case. In (a) and (b) both, sticking probability was taken as unity.*

**Fig.6.** *(a) Variations of fractal dimension with Monte Carlo step for different sticking probabilities in light water hydration case. (b) Variations of fractal dimension with Monte Carlo step for different sticking probabilities in heavy water hydration case.*

**Fig.7.** *(a) Variations of square of characteristic length with Monte Carlo step for different sticking probabilities in light water hydration case. (b) Variations of square of characteristic length with Monte Carlo step for different sticking probabilities in heavy water hydration.*

**Fig.8.** *(a) Evolutions of the simulated scattering profile with Monte Carlo steps at different q values for light water hydration case. (In inset) Evolution of simulated scattering intensity with Monte Carlo steps at q~0 for light water hydration case. (b) Evolutions of the simulated scattering profile with Monte Carlo steps at different q values for heavy water hydration case. (In inset) Evolution of simulated scattering intensity with Monte Carlo steps at q ~ 0 for heavy water hydration case.*

**Fig.9.** *(a) Initial DLA cluster. (b)-(e) Evolution of DLA cluster while hydrated with light water(on left) and heavy water(on right) at different MC step indicated in the figure.*

**Fig.10.** *Fitting of $D_f$ vs. number of MC steps for light water hydration of cement as in Fig.4.(a) with Eq.(10) in the proposed growth model for $L\sim0.26$, $N_{ci}\sim6.72\times10^2$, $\tilde{\eta}\sim2.45$, $N_{cs}(0)\sim3.83\times10^3$, $c\sim1.00\times10^{-5}$, $K_{c\rightarrow g}\sim3.81\times10^{-8}$ min.$^{-1}$. This is not a unique set of values. It varies depending on the guessed values provided before the fitting starts, which implies that the same temporal variation of $D_f$ is possible for clusters of different morphology.*

**Fig.11.** *Variation of $D_f$ with number of MC steps and real time of hydration in the case of light water hydration. In the dotted sections, $D_f$ changes from 2.25 to 2.3 for both cases and the*

*corresponding change in number of MC steps is (b-a)=5800 MC steps and in real time is (B-A)=30 min. (b-a)/(B-A)~193 MC steps/min i.e. the change that 193 MC steps bring about into the simulated system, occurs in 1 min. in the real system. But this factor varies depending on the state of the system in different time regimes.*

**Fig.12.** *$N_g(t)$ vs. t and $N_G(t)$ vs. t as proposed in growth model with $\tilde{\eta}\sim 2.45$, $N_{cs}(0)\sim 3.83\times 10^3$, $c\sim 1.00\times 10^{-5}$, $K_{c\to g}\sim 4.85\times 10^{-3}$min.$^{-1}$, $K_{g\to G}\sim 3.35\times 10^{-4}$min.$^{-1}$, $\xi\sim 2.25$. This is not a unique set of values for the reason already mentioned in Fig.10.*

**Fig.13.** *$D_f$ vs. t as proposed in growth model with $L\sim 0.31$, $N_{ci}\sim 7.20\times 10^2$, $\tilde{\eta}\sim 2.45$, $N_{cs}(0)\sim 3.83\times 10^3$, $c\sim 1.00\times 10^{-5}$, $K_{c\to g}\sim 4.85\times 10^{-3}$min$^{-1}$, $K_{g\to G}\sim 3.35\times 10^{-4}$min$^{-1}$, $\xi\sim 2.25$ (top). $D_f$ vs. number of MC steps for heavy water hydration of cement by computer model (bottom). This is not a unique set of values for the reason already mentioned in Fig.10.*